\documentclass[conference]{IEEEtran}
\IEEEoverridecommandlockouts

\usepackage{cite}
\usepackage{amsmath,amssymb,amsfonts}
\usepackage{graphicx}
\usepackage{textcomp}
\usepackage{xcolor}
\usepackage[linesnumbered,ruled,vlined]{algorithm2e}

\usepackage{booktabs}
\usepackage{multirow}
\usepackage{bbding}
\usepackage{pifont}
\usepackage{array}
\usepackage{bm}
\usepackage{color,colortbl}
\usepackage{xcolor}
\definecolor{gautamblue}{RGB}{215, 238, 247}
\definecolor{gautamred}{RGB}{255, 186, 184}
\definecolor{gautamgreen}{RGB}{217,255,219}
\definecolor{gautamorange}{RGB}{255, 219, 153}
\definecolor{gautamviolet}{RGB}{224, 220, 244}
\def\HiLiBlue{\leavevmode\rlap{\hbox to \hsize{\color{gautamblue}\leaders\hrule height .8\baselineskip depth .5ex\hfill}}}
\def\HiLiYellow{\leavevmode\rlap{\hbox to \hsize{\color{gautamorange}\leaders\hrule height .8\baselineskip depth .5ex\hfill}}}
\def\HiLiGreen{\leavevmode\rlap{\hbox to \hsize{\color{gautamgreen}\leaders\hrule height .8\baselineskip depth .5ex\hfill}}}
\def\HiLiRed{\leavevmode\rlap{\hbox to \hsize{\color{gautamred}\leaders\hrule height .8\baselineskip depth .5ex\hfill}}}

\usepackage{filecontents}
\usepackage{subcaption}

\usepackage{tikz}
\usepackage{amsmath}
\usepackage{amssymb}
\usepackage{enumitem}
\setlist{nosep}
\usepackage{url}

\begin{document}
\title{Exploiting Dependency-Aware Priority Adjustment for Mixed-Criticality TSN Flow Scheduling
\thanks{This study is supported by the National Key R\&D Program of China under Grant 2022YFE0112600, the National Natural Science Foundation of China under Grants No. (U1909207, 62302439, U23A20296), and the Fundamental Research Funds for the Central Universities (226-2024-00004). \textit{Chaojie Gu is the corresponding author}.}}

\author{\IEEEauthorblockN{Miao Guo\IEEEauthorrefmark{2}, Yifei Sun\IEEEauthorrefmark{2}, Chaojie Gu\IEEEauthorrefmark{2}, Shibo He\IEEEauthorrefmark{2}, Zhiguo Shi\IEEEauthorrefmark{3}}
\IEEEauthorblockA{\IEEEauthorrefmark{2}\textit{College of Control Science and Engineering, Zhejiang University, Hangzhou, China} \\
\IEEEauthorrefmark{3}\textit{College of Information Science and Electronic Engineering, Zhejiang University, Hangzhou, China} \\
\{gm\_oct, 22232080, gucj, s18he, shizg\}@zju.edu.cn}

}

\maketitle

\begin{abstract}
Time-Sensitive Networking (TSN) serves as a one-size-fits-all solution for mixed-criticality communication, in which flow scheduling is vital to guarantee real-time transmissions. 
Traditional approaches statically assign priorities to flows based on their associated applications, resulting in significant queuing delays.
In this paper, we observe that assigning different priorities to a flow leads to varying delays {due to different shaping mechanisms applied to different flow types.
Leveraging this insight,} we introduce a new scheduling method {in mixed-criticality TSN} that incorporates a priority adjustment scheme among diverse flow types to mitigate queuing delays and enhance schedulability.   
Specifically, we propose dependency-aware priority adjustment algorithms tailored to different link-overlapping conditions. 
Experiments in various settings validate the effectiveness of the proposed method, which enhances the schedulability by 20.57\% compared with the SOTA method.
\end{abstract}

\section{Introduction}
\label{sec:introduction}
Time-Sensitive Networking (TSN) has arisen as a promising technology in automotive\cite{9799732} and industrial automation\cite{10390539}. Enabling the convergence of Information Technology (IT) and Operational Technology (OT), TSN serves as a one-size-fits-all solution to cope with the mixed-criticality communication among drastically increasing applications\cite{8695835}. 

In TSN, the mixed-criticality traffic is divided into three priority classes with different Quality-of-Service (QoS) requirements: Time-Triggered (TT) traffic with tight timing constraints, Audio-Video Bridging (AVB) traffic with bounded latency but less stringent requirements, and Best-Effort (BE) traffic with no timing guarantees\cite{8402374}. To satisfy different QoS requirements, TSN specifies various mechanisms {implemented in the switches} for all types of traffic: Time-Aware Shaper (TAS) for queue isolation and deterministic transmission of TT traffic\cite{standardQbv}, Credit-Based Shaper (CBS) for rate-limiting and burst-free delivery\cite{standardBA} of AVB traffic, and preemption mechanism for high bandwidth utilization of mixed traffic\cite{10212759}. 

Based on the applied mechanisms, traffic scheduling performs admission control and generates the routing and timing behavior for data flows, which is of vital importance to guarantee deterministic and real-time transmission in TSN. Thus, many research works focus on investigating TSN traffic scheduling\cite{9709655,LIN2021102141}.
However, existing studies mainly focus on the scheduling of TT flows in the TAS mechanism, unapplicable to mixed-criticality scenarios {with the presence of all the mechanisms}. 

\begin{figure}[t]
\centering
\begin{subfigure}[t]{.25\textwidth}
\centering
\includegraphics[width=\linewidth]{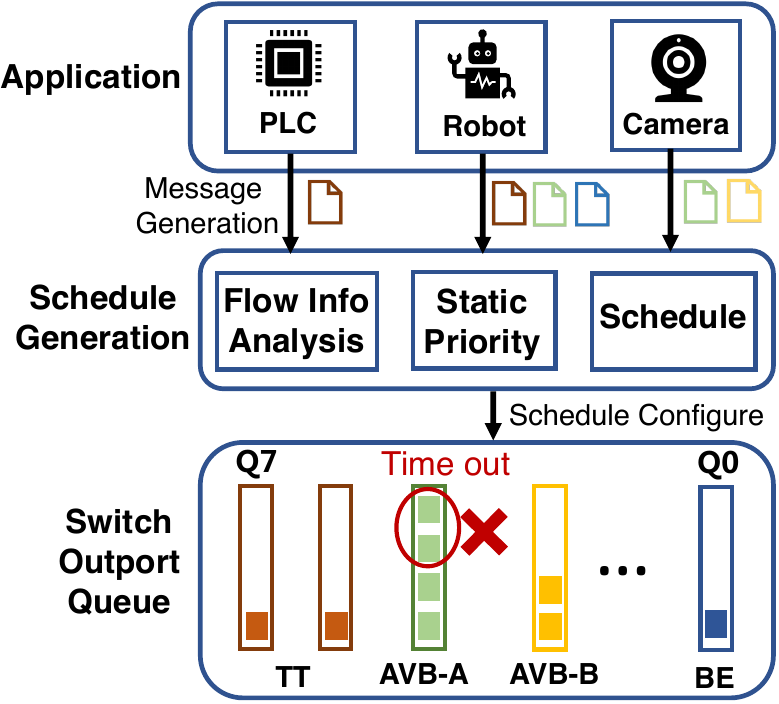}
\caption{}
\label{fig:intro1}
\end{subfigure}
\hfill
\begin{subfigure}[t]{.19\textwidth}
\centering
\includegraphics[width=\linewidth]{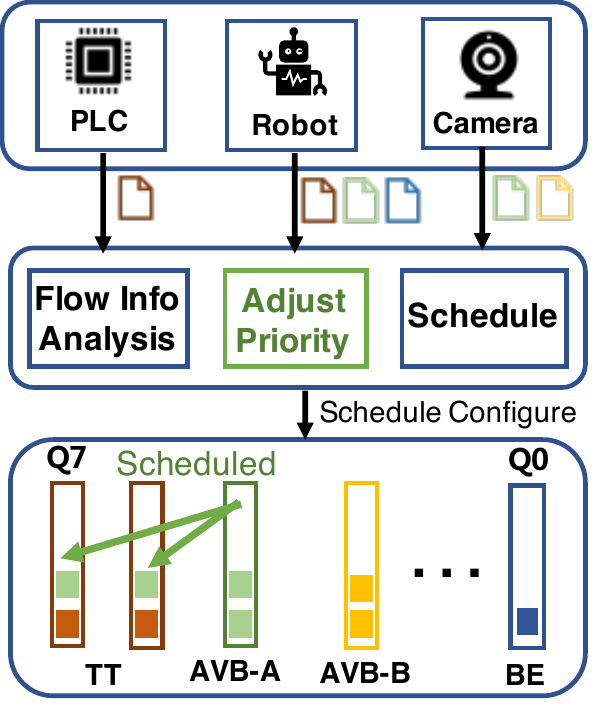}
\caption{}
\label{fig:intro2}
\end{subfigure}
\caption{Comparison between (a) Static priority assignment and (b) Adaptive priority adjustment. In (a), AVB-A flows experience severe queuing delay and eventually reach a timeout state. After adaptive priority adjustment in (b), the delays of tightly time-constrained flows are reduced. All flows are scheduled.}
\label{fig:intro}
\end{figure}

{In mixed-criticality scenarios, flows are typically categorized based on the applications they are associated with.} For example, the Control Data from Industrial Automation and Control System (IACS) applications are mapped to TT traffic\cite{8458130}. Voice messages requiring latency and jitter lower than 10ms from Audio/Video applications are mapped to AVB-A (detailed in Sec.~\ref{sec:flow model}) traffic\cite{8458130}. 
{After the assignment of the traffic type, the corresponding flow will be cached in the predefined outport queue on the switch.}
{According to IEEE 802.1Qbv standard, among the eight queues on the output port of each switch, one or more are for TT flows, two are for AVB flows, and the remaining are for BE flows.}
{Consequently, the static assignment of flow types may cause a domino effect of delays of a particular type of flow when the outport queue overflows.}
As Fig.~\ref{fig:intro1} shows, in the static priority assignment case, the flows of AVB-A traffic type accumulate a lot due to a large number of corresponding application messages, which causes severe queuing delay and even timeout of AVB-A flows. 

{In practice, different applications have different QoS requirements and the flow priorities are typically not evenly distributed\cite{8458130}. If we adopt the static priority assignment strategy, different traffic types will experience diverse delay conditions.}
Some may experience severe queuing delays while others barely accumulate, which leaves free space to adjust the priority of congested flows to alleviate heavy congestion. 
Meanwhile, we observe that a flow could experience different delays when assigning different traffic types to it due to different shaping mechanisms. This leaves an opportunity to reduce the delays of congested flows within their deadlines and enhance schedulability by adjusting their priorities. Thus, we propose an adaptive priority adjustment scheme among different traffic types for mixed-criticality traffic scheduling.
As Fig.~\ref{fig:intro2} shows, we adjust part of the AVB-A flows to TT priority to mitigate the queuing delay from the same-type flows, which reduces the delay of AVB-A flows and enhances the schedulability.  

In this paper, we present the adaptive priority adjustment scheme among different traffic types for mixed-criticality traffic scheduling, which targets flows for multiple types in the presence of TAS, CBS, and preemption mechanisms. We compare it with other SOTA methods. The results from experiments in realistic topologies show that we acquire a 20.57\%/29.68\% schedulability improvement over Tabu/Static methods. The contributions are as follows:

\begin{itemize}
    \item We introduce the first mixed-criticality traffic scheduling method globally solving the priority adjustment problem among different traffic types with the presence of TAS, CBS, and preemption mechanisms. 

    \item We propose dependency-aware priority adjustment algorithms for flows in diverse link-overlapping conditions, which can flexibly determine the adjusting policy with the knowledge of both network conflict and flow features. 
    \item We conduct experiments on three typical realistic topologies and different flow volumes to validate the effectiveness and adaptiveness of the proposed method. 
\end{itemize}

\section{System Model}
\label{sec:system model}
\subsection{Network Model}
The topology of the TSN network is modeled as a directed graph $G=\{V, E\}$. The set of vertices $V$ is composed of the set of end devices $ES$ and the set of switches $SW$, denoted as $V=ES\cup SW$. End devices serve as the source and destination of flows while switches only forward flows. The set of edges $E$ represents data links. The network supports full-duplex transmission, thus the links are directed. For example, two vertices $v_a, v_b \in E$ determine two links $[v_a, v_b]$ and $[v_b, v_a]$. The transmission rate of each link $l$ is considered constant, denoted as $l.s$.
\subsection{Flow Model}
\label{sec:flow model}
According to diverse QoS requirements of data transmissions, TSN flows are categorized into three priority classes: TT flows, AVB flows (Class A and B), BE flows. Their priorities ranked from high to low are: TT $>$ Class A $>$ Class B $>$ BE.

The TSN flow can be modeled as a 6-tuple consisting of source, destination, period, frame length, maximum end-to-end delay (deadline), and priority: 
\begin{align}
    &{\forall}f_{i} {\in} F, ~i {\in} [0, ~n-1],  \notag \\
    &f_{i}=\{src,dst,period,size,ddl,prio\},
\end{align}
where $F$ is the set containing all the flows and $n$ is the flow number. $F$ can be divided into three sub-sets according to the classification, denoted as $F=F^{TT}\cup F^{AVB}\cup F^{BE}$.

Since the flow is periodic, the k-th periodic occurrence of a flow is called the k-th instance. For $f$ starting from $v_a$ and ending at $v_b$, its route $f.R$ is an ordered sequence $[[v_a,v_{a+1}],...,[v_{b-1},v_b]]$. Moreover, the transmission time of $f$ on link $l$ is $f.C=\frac{f.size}{l.s}$. In this paper, we assume all flows have the size of one MTU (Maximum Transmission Unit), i.e., 1500 bytes. If a flow $f$ can be successfully accommodated in the network, we denote it as $S(f_i)=1$. Otherwise, $S(f_i)=0$.

For each TT flow, we need to generate the departure time (offset) $\phi$ on each link along its route due to its determinism property. The offset sequence of TT flow $f$ is denoted as $f.O=[f^{[v_a,v_{a+1}]}.\phi,...,f^{[v_{b-1},v_{b}]}.\phi]$. 
\subsection{Switch Model}
\begin{figure}[t]
\centering
\includegraphics[width=.7\linewidth]{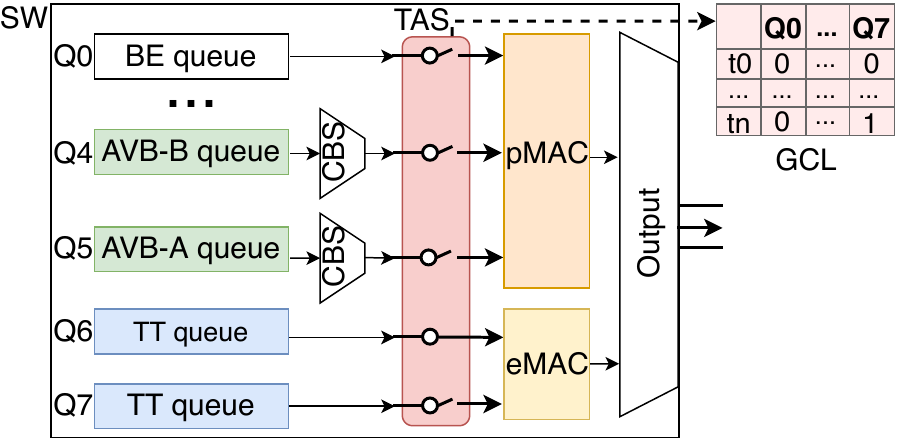}
\caption{The TSN switch model supporting preemption mode. Express Mac (eMac) and preemptable Mac (pMac) are two Mac types defined in the TSN preemption specification. Flows in eMac can preempt flows in pMac.} 
\label{fig:switch model}
\vspace{-1em}
\end{figure}
As is shown in Fig.~\ref{fig:switch model}, the egress port in the TSN switch contains eight queues for storing and forwarding frames. Among the queues, two are TT queues, two are AVB queues, and the remaining are BE queues. 
Flows of the three types are isolated spacially with the corresponding queues.

During the transmissions of mixed-criticality flows, different forwarding and shaping mechanisms are applied to them.

\noindent\textbf{For TT flows:} IEEE 802.1Qbv\cite{standardQbv} specifies TAS via the insertion of a gate control list (GCL) to define the exact queue transmission times of frames on each egress port.

\noindent\textbf{For AVB flows:} In order to prevent the starvation of lower priority messages,
IEEE 802.1BA\cite{standardBA} specifies CBS to regulate the transmission of AVB-A and AVB-B flows.
Each AVB class has an associate credit parameter. Whenever there is a pending message in the queue of a flow class, the transmission can occur only if the associated credit is zero or higher. 

\noindent\textbf{Preemption in mixed-criticality transmission:} 
The transmission of an AVB and BE frame can be interrupted by the transmission of a TT frame and resumed once the TT frame is fully transmitted. However, the preemptable flow is allowed to continue transmitting up to 123 bytes even after the transmission gate of the TT flow has opened. After preemption, the preemptable flow is resumed with a new header, which should account for frame overhead and additional delay. 

\section{Problem Formulation}
\label{sec:problem formulation}
\begin{figure}[t]
\centering
\includegraphics[width=.7\linewidth]{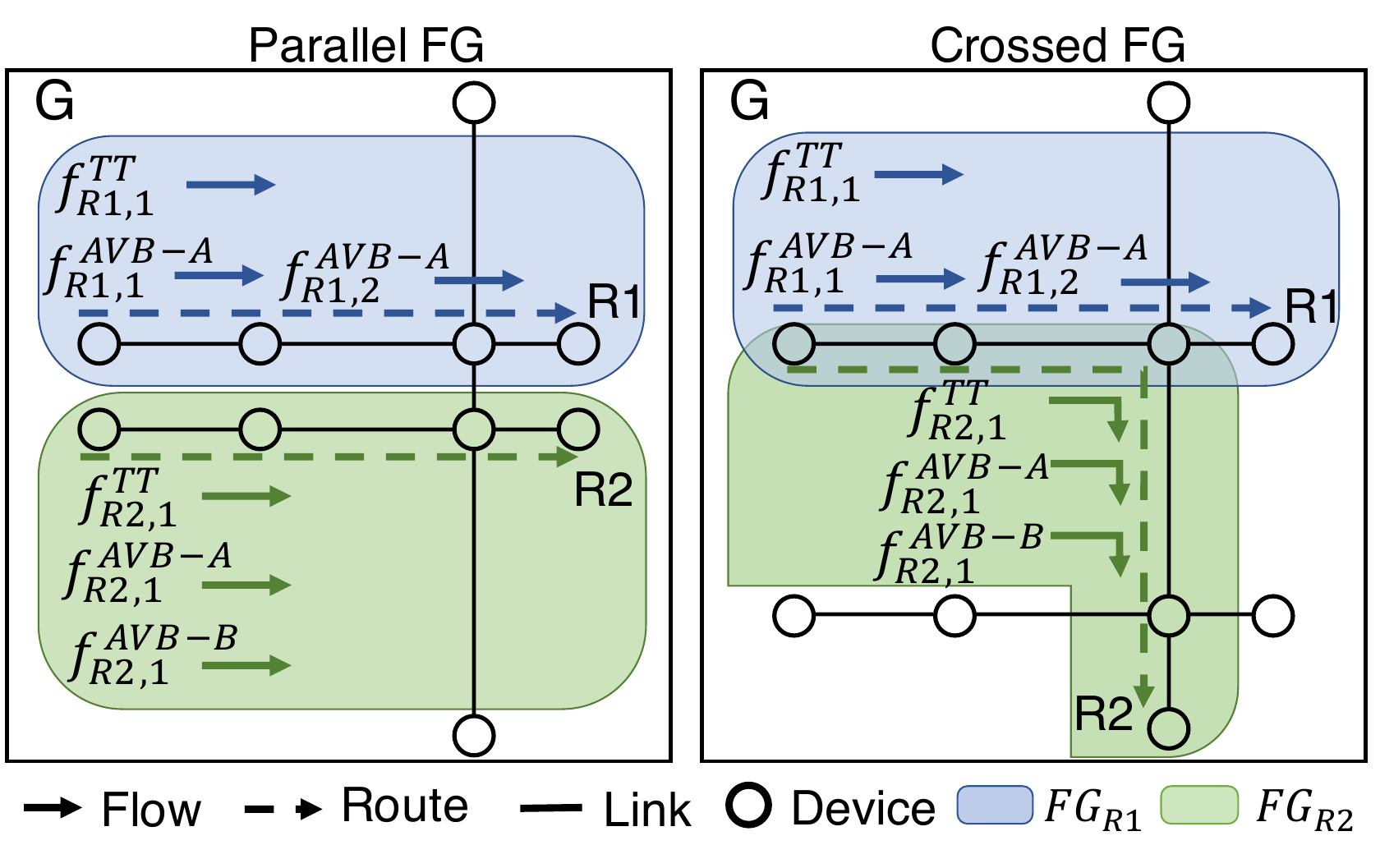}
\caption{Parallel FGs and crossed FGs. Flow $f_{Ri,j}^{x}$ means the $j$th flow of class $x$ in $FG_{Ri}$.}
\label{fig:parallel_cross_fg}
\vspace{-1em}
\end{figure}
To accommodate mixed-criticality flows in TSN, the worst-case delay (WCD) of each flow should be within the maximum end-to-end delay. To work out the WCD of flows, we use RTA as a timing analysis method. The WCD of flow $f$ is denoted as $f.RT$. The response time (RT) of flows in this paper follows the analysis in \cite{LOBELLO2020153}. 

To reduce the queuing delay and increase the schedulability, traffic type adjustment plays a vital role as flows with different traffic types adopt different shaping policies and thus generate different RTs. As a result, the decision variables of the problem include the adjusted priorities of all the flows (apart from BE flows) $F.prio$ to replace the static priority $F.prio_{static}$ and the schedules of TT flows $F^{TT}.O$. We do not adjust the priority of BE flows since it has no timing requirements.

To solve the problem, the following constraints should be satisfied.

\noindent\textbf{Deadline Constraint}:
The worst-case delay of each flow should be within its deadline. $\forall f \in F, f.RT\le f.ddl$.

\noindent\textbf{Priority Constraint}:
For statically defined TT flows, their priorities should not be adjusted lower than that due to their deterministic requirement. $\forall f \in F^{TT}, f.prio=f.prio_{static}$.

\noindent\textbf{TT Transmission Constraint}:
{To calculate the transmission windows of TT flows, we follow the constraints elaborated in \cite{craciunas2016scheduling}.} (1) Frame Constraint: The offset of any flow has to be greater than or equal to 0 and the transmission window has to fit within the period. (2) Flow Transmission Constraint: The offsets of a flow must follow the sequential order along the route. (3) Link Constraint: Two flows routed through the same link can not overlap in the time domain. (4) Deterministic Queue Constraint: Packets of link-sharing flows either enter different queues or are scheduled so their arrival times are far apart to avoid interleaving.

\noindent\textbf{Optimization Objective:} The optimization objective is maximizing the number of successfully scheduled flows: ${Maximize} \sum_{f_i\in F}S(i)$.

\section{Optimization Strategy}
\label{sec:optimization strategy}

The priority adjustment problem associated with the TT scheduling problem has been proved to be NP-hard\cite{LIN2021102141}, meaning there is no polynomial-time algorithm that can generate the optimal solution. 

Hence, by exploiting insights from both the problem and response time analysis, we propose effective algorithms that can strike a balance between results quality and efficiency. 
{Since adjusting the priority of flows would affect the schedulability of other flows sharing the same link, there are dependencies among flows in different link-overlapping conditions. To clarify the dependencies, we first define \textit{flow group (FG)} as follows:}
\begin{equation}
    FG_{R1}=\{f|f\in F,f.R=R1,R1\in G\}, 
\end{equation}
which is a set of flows sharing the same route in network $G$. The shared route of the flow group is denoted as $FG_{R1}.R$. As Fig.~\ref{fig:parallel_cross_fg} shows, when $FG_{R1}.R\cap FG_{R2}.R=\emptyset$, the two FGs are parallel, otherwise, they are crossed. 
According to the two dependency cases, we divide the investigation into two cases.

\subsection{Parallel-FG case}
\label{sec:parallel-fg}
\begin{figure}[t]
\centering
\includegraphics[width=.8\linewidth]{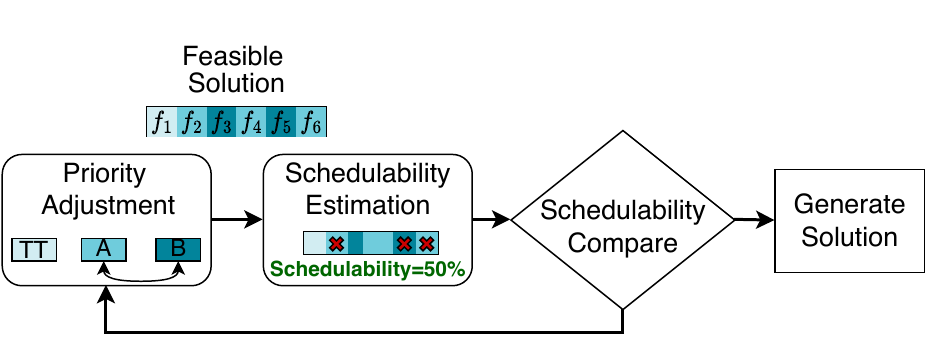}
\caption{Scheduling process of AVB-A and AVB-B adjustment. The priority adjustment determines the priority for all the AVB flows, which forms the feasible solution.} 
\label{fig:schedule_process}
\vspace{-1em}
\end{figure}
In this case, all the FGs are parallel, meaning the routes of different FGs share no links. Since RTA considers flows sharing the same link, in the parallel-FG case, the RTA of each flow needs to consider only the flows within one FG and thus is independent of other FGs. For example, in the parallel-FG case in Fig.~\ref{fig:parallel_cross_fg}, the RT computation of flow $f_{R1,2}^{AVB-A}$ is only related to other flows in $FG_{R1}$ rather than $FG_{R2}$ because of no link sharing. Therefore, we could solve the problem of each FG in parallel due to the interdependence.

Within each FG, we take two steps to adjust the flow priorities. Considering the preemption and delay interference from higher-priority flows to lower-priority flows, we take the following priority-descending order. 

\noindent\textbf{TT adjustment phase:} We first schedule the statically defined TT flows to satisfy the \textit{priority constraint}. To guarantee low latency of flows with tight deadlines, we sort AVB flows in deadline-ascending order and incrementally schedule them as TT flows. When any AVB flows fail to be scheduled as TT flows, their priorities remain unchanged.

\noindent\textbf{AVB-A and AVB-B adjustment phase:} 
To allocate two AVB priorities to flows properly,
we first assign AVB-A to all the flows and then adjust flows to AVB-B in the deadline-descending order until the current-best solution occurs. 
The scheduling process of {this phase} follows Fig.~\ref{fig:schedule_process}. Each adjustment determines a feasible solution. By estimating and comparing the schedulability of different solutions, we obtain the final result with the highest schedulability. However, different estimation policies could influence the schedulability. 

To enhance the schedulability of the solution, we design Alg.~\ref{code:get num} to estimate the schedulability of AVB flows.

\begin{algorithm}[t]
    \SetAlgoLined
    \SetKwFor{For}{for}{do}{end for}
    \footnotesize
    \KwIn{$F^{AVB-x}$ of each FG\textcolor{gray}{~// x=A or B\\}}
    \KwOut{the maximum number of successfully scheduled flows n}
    Flist~$\leftarrow$ sort $F^{AVB-x}$ in deadline ascending order,~n=0\\
    \While{True}{
    n=Num(Flist)\\
    \eIf{$Flist[0].ddl \ge RT(n)$}{
    Flist.schedulable=True, ~break\\
    }
    {
    Flist[0].schedulable=False,~Delete Flist[0] from Flist
    }
    }
    \Return n
    \caption{Schedulability estimation $S()$---get the number of scheduled AVB-x flows in parallel-FG case.\label{code:get num}}
\end{algorithm}
Under a specific assignment solution, we first assume the status of all the flows as scheduled and traverse the AVB-x set in deadline ascending order (line 1) to find the schedulable flow with the smallest deadline that could determine the maximum number $n$ of scheduled flows in lines 4-8. $RT(n)$ represents the RT of the remaining n AVB-x flows.
Subsequently, we conclude the scheduling process of the parallel-FG case in Alg.~\ref{code:parallelFG}, where the function of Alg.~\ref{code:get num} is denoted as \textit{S()}. {The time complexity of Alg.~\ref{code:get num} is $O(n^2)$. $n$ is the number of $F^{AVB-x}$.}

Specifically, the scheduling of TT flows adopts the ASAP (As Soon As Possible) algorithm\cite{Raagaard2018OptimizationAF} in line 2, which allocates the earliest feasible time to TT flows on each hop. In lines 4-9, we conduct TT adjustment by sorting AVB flows in deadline-ascending order and incrementally scheduling them as TT flows. In lines 11-22, we conduct AVB-A and AVB-B adjustments by first assigning AVB-A to all the flows and then adjusting flows to AVB-B in the deadline-descending order as flows with large deadlines have more time margin for queuing and waiting as lower-priority AVB-B flows. The above process repeats until the current-best solution occurs. {The time complexity of Alg.~\ref{code:parallelFG} is $O(n^3+n_{AVB})$. $n$ and $n_{AVB}$ are the numbers of all the flows and AVB flows, respectively.}

\begin{algorithm}[t]
    \SetAlgoLined
    \SetKwFor{For}{for}{do}{end for}
    \footnotesize
    \KwIn{Static-priority assigned flow set $F$ of each FG}
    \KwOut{Adjusted priority $F.prio$, schedule plan of TT flows $\Phi$}
    \textcolor{gray}{// TT adjustment\\}
    $\Phi$=ASAP($F^{TT}$)\\
    Sort $F^{AVB}$ in deadline ascending order\\
    \For{$f\in F^{AVB}$}{
    Schedulable,$\Phi$=ASAP(f,$\Phi$)\\
    \If{$Schedulable==True$}{
    f.prio=TT,~f$\rightarrow F^{TT}$
    }
    }
    \textcolor{gray}{// AVB-A and AVB-B adjustment\\}
    $F^{AVB}.prio=AVB-A,~F^{AVB-A}=F^{AVB},~F^{AVB-B}=\emptyset$\\
    $N_{best}=S(F^{AVB-A})$\\
    Sort $F^{AVB-A}$ in deadline ascending order\\
    \For{$f \in Reversed (F^{AVB-A})$}{
    $f.prio=AVB-B,~f\rightarrow F^{AVB-B}$\\
    $N_{cur}=S(F^{AVB-A})+S(F^{AVB-B})$\\
    
    \eIf{$N_{cur} \ge N_{best}$}{
    $N_{best}=N_{cur},~Solution=F.prio$\\
    }
    {
    break \\
    }
    }
    \Return Solution,~$\Phi$
    \caption{Schedule flows in the parallel-FG case.\label{code:parallelFG}}
\end{algorithm}

\subsection{Cross-FG case}
The challenge of the cross-FG case is the adjustment in one FG could influence the RTs of flows in other FGs due to the link sharing among FGs, affecting the overall schedulability. Therefore, we should jointly consider the \textit{link overlap} among FGs and the original \textit{flow features} to decide the priority adjustment. 

Similar to the adjusting steps in Sec.~\ref{sec:parallel-fg}, we jointly consider both aspects in the following two phases.

\noindent\textbf{TT adjustment phase:} We first schedule the statically defined TT flows and then adjust part of the AVB flows to TT. When determining the adjustment order, both link overlap and flow features should be considered.
Link overlap can be qualified as $TTConfliction$. For the AVB flow to be adjusted to TT, $TTConfliction$ is defined as the number of all the TT flow instances on its route. 
\begin{equation}
    f.TTConfliction=\sum_{l\in f.R}\sum_{f_j\in F^{TT} \land l\in f_j.R}\frac{T_{sched}}{f_j.period},
\end{equation}
where $T_{sched}$ is the hyper period, i.e., the least common multiple (LCM) of all the periods of TT flows.
A large $TTConfliction$ means a large confliction possibility among TT flows when adjusting the flow as TT. To avoid conflict, we give flows with small $TTConfliction$ the precedence for adjusting to TT. 

Combining link overlap and flow feature, we design a metric $TTAdjust$ to define the precedence of AVB flows for TT adjustment since small deadlines mean critical emergencies and flows with large deadlines have more time margin for queuing and waiting as low-priority AVB flows. 
\begin{equation}
    f.TTAdjust=\frac{1}{f.TTConfliction\times f.ddl}.
\end{equation}
Flows with large $TTAdjust$ are prioritized for TT adjustment.

\begin{figure}[t]
\centering
\begin{subfigure}[t]{.17\textwidth}
\centering
\includegraphics[width=\linewidth]{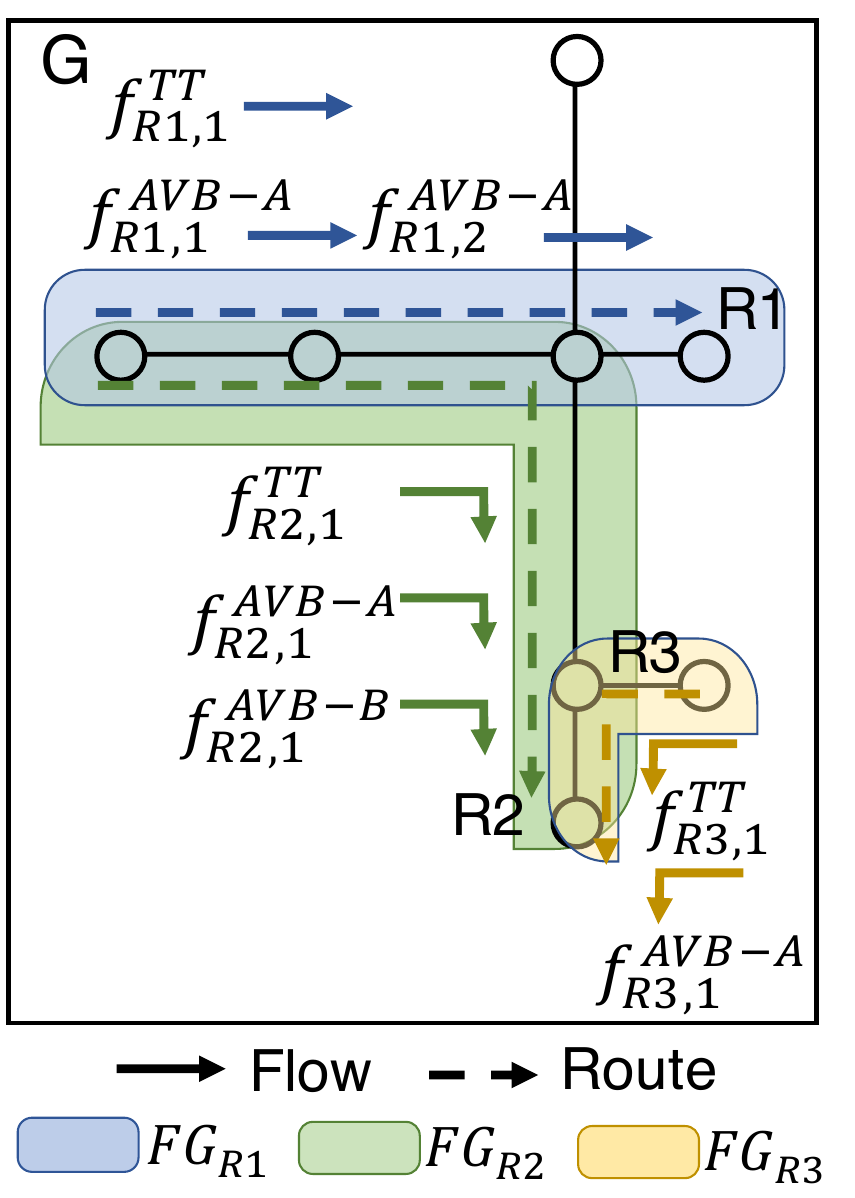}
\caption{}
\label{fig:3FG}
\end{subfigure}
\hfill
\begin{subfigure}[t]{.21\textwidth}
\centering
\includegraphics[width=\linewidth]{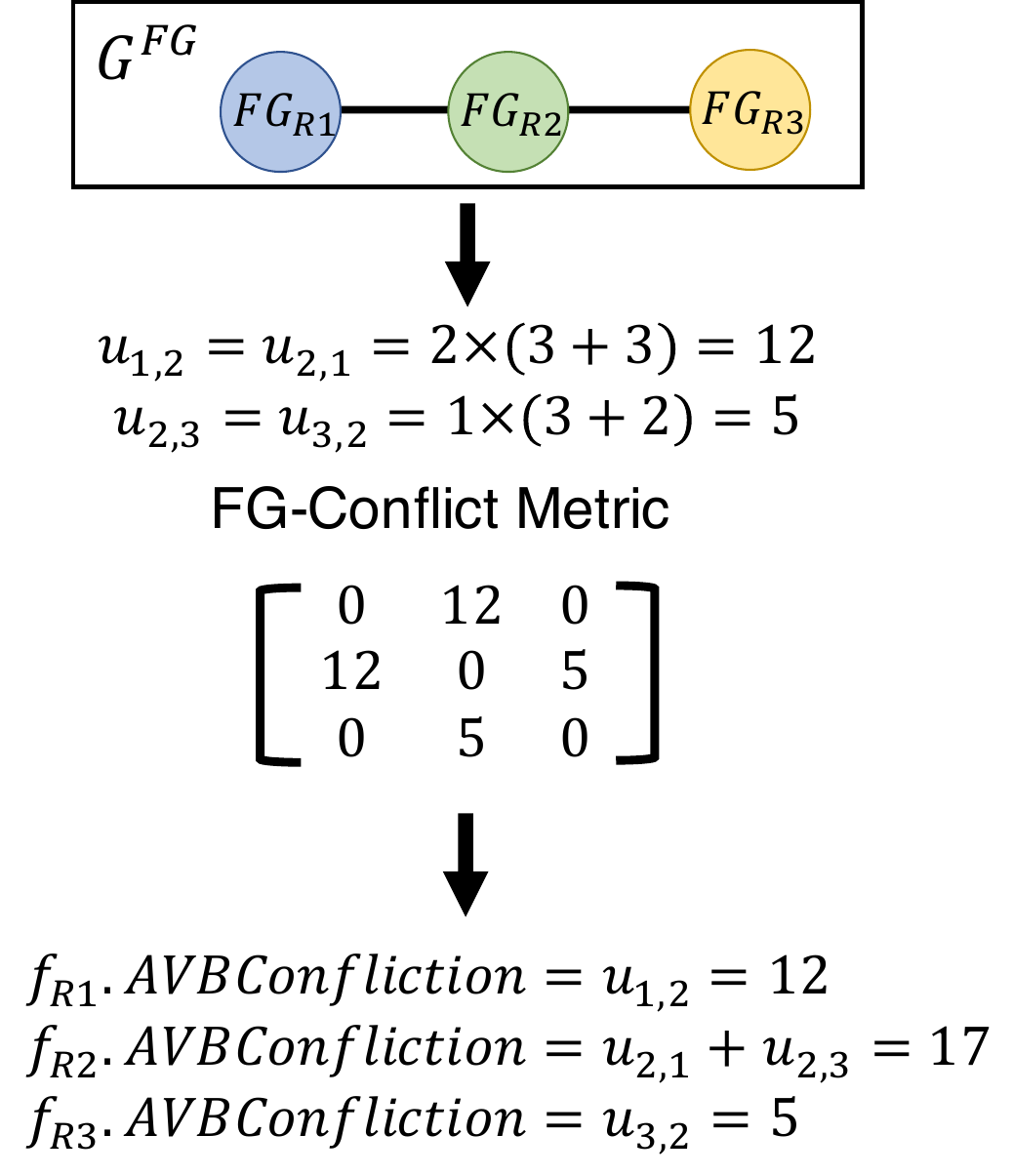}
\caption{}
\label{fig:FGgraph}
\end{subfigure}
\caption{Illustration of FG-conflict graph. (a) The network topology and flows. $f_{Ri,j}^{x}$ means the $j$th flow of class $x$ in $FG_{Ri}$. (b) The FG-conflict graph of (a). $u_{i,j}$ is the edge weight between $FG_{Ri}$ and $FG_{Rj}$. $f_{Ri}$ means the flow in $FG_{Ri}$.}
\label{fig:FG-conflict graph}
\vspace{-1em}
\end{figure}

\noindent\textbf{AVB-A and AVB-B adjustment phase:} We first assign AVB-A to all the remaining AVB flows and then adjust part of the flows to AVB-B. When determining the adjusting order, link overlaps and flow features should be considered. Link overlap can be qualified
as $AVBConfliction$. To depict the congestion degree of AVB flows, we construct an undirected weighted graph $G^{FG}=\{\Gamma,~U\}$ named FG-conflict graph. In $G^{FG}$, each FG $\gamma_i \in \Gamma$ represents a node, and the FG-conflict metric $u_{i,j}= u_{j,i} \in U$ represents the edge weight between FG nodes $\gamma_i$ and $\gamma_j$, which is defined as follows:
\begin{equation}
    u_{i,j}=|FG_i.R \cap FG_j.R| \times (|FG_i|+|FG_j|),
\end{equation}
where $|FG_i.R \cap FG_j.R|$ is the number of shared links between $FG_i$ and $FG_j$, and $|FG_i|$ is the total number of successfully scheduled TT flows and the AVB flows to be scheduled in $FG_i$. It takes both link overlap and flow congestion into account and can represent the dependency between two FGs. Based on it, the $AVBConfliction$ of flow $f$ is defined as:
\begin{equation}
    f.AVBConfliction=\sum_{u_{i,j}\ne 0} u_{i,j}, ~f\in FG_i,
\end{equation}
representing the weighted degree of the FG node to which $f$ belongs. 
To illustrate it, we take Fig.~\ref{fig:FG-conflict graph} as an example. 

Combining link overlap and flow feature, we design a metric $AVBAdjust$ to define the precedence of AVB adjustment:
\begin{equation}
    f.AVBAdjust=\frac{f.AVBConfliction}{f.ddl}.
\end{equation}

Large $AVBAdjust$ means more resource conflict and tight deadlines. Flows with small $AVBAdjust$ are prioritized for adjusting to AVB-B since they have large resource and time margins for queuing as lower-priority flows.

{The scheduling algorithm in the cross-FG case is similar to Alg.~\ref{code:parallelFG}, apart from the following differences. (1) Instead of regarding each FG as input, we input the whole flow set $F$ due to the cross-FG dependencies. (2) In line 3, we sort $F^{AVB}$ in $TTAdjust$ descending order. (3) In line 13, we sort $AVBAdjust$ in descending order. (4) In line 20, we restore the adjusted flow to $F^{AVB-A}$ in the original order and proceed with the loop. Moreover, we change the schedulability estimation policy $S()$ into sorting flows in $AVBAdjust$ descending order to check the RT of each flow. }

\section{Evaluation}
\label{sec:evaluation}
\subsection{Experimental setup}
\subsubsection{Baseline}
To demonstrate the effectiveness of the proposed algorithm, we compare it with four algorithms. 1. \textit{The static algorithm (Static)} assigns priorities to flows based on the original setting. 2. \textit{The upgrade algorithm (Upgrade)}\cite{9709655} conducts priority compensation greedily based on the original priority. 3. \textit{The tabu algorithm (Tabu)}\cite{10.1145/3371708} adjusts the priority with tabu heuristic. 4. \textit{The GRASP algorithm (GRASP)}\cite{8548553} adjusts the priority with GRASP heuristic.

\subsubsection{Network setting}
To test the scalability and robustness of the proposed algorithm, we conduct experiments on three realistic topologies: SAE (Society of Automotive Engineers)\cite{10.1145/3371708}, AFDX (Avionics Full-Duplex
switched Ethernet)\cite{9522239}, and Ladder\cite{9522239}. 

\subsection{Experimental Results}
\subsubsection{Performance in different traffic volumes}

\begin{figure*}[t]
\centering
\begin{subfigure}[t]{.27\linewidth}
\centering
\includegraphics[width=\linewidth]{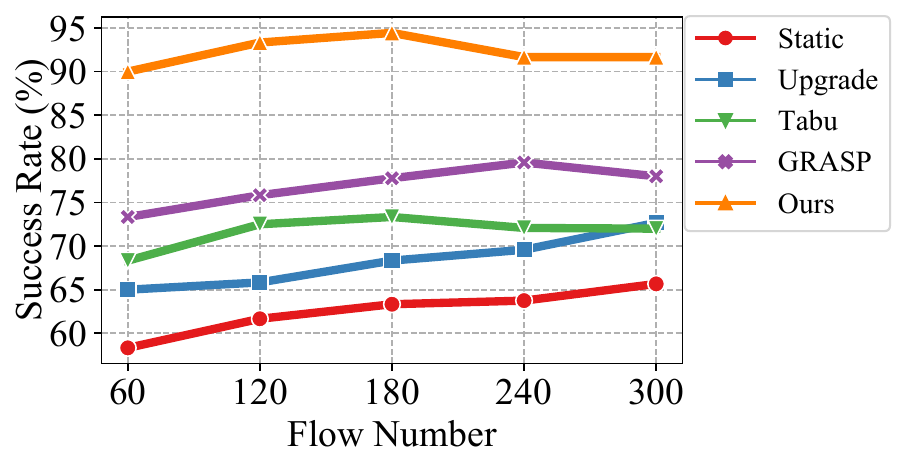}
\caption{Scheduling success rate.}
\label{fig:sucrate_flownum}
\end{subfigure}
\hfill
\begin{subfigure}[t]{.27\linewidth}
\centering
\includegraphics[width=\linewidth]{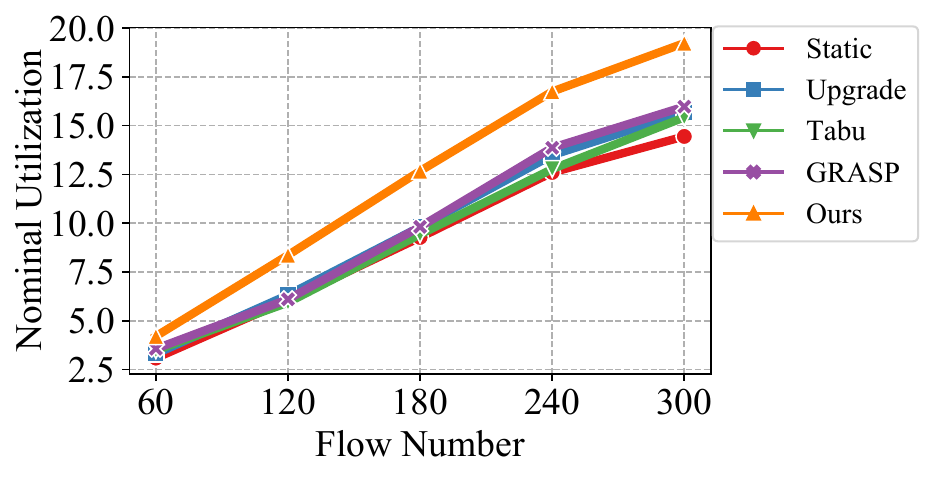}
\caption{Nominal Utilization.}
\label{fig:resuti_flownum}
\end{subfigure}
\hfill
\begin{subfigure}[t]{.27\linewidth}
\centering
\includegraphics[width=\linewidth]{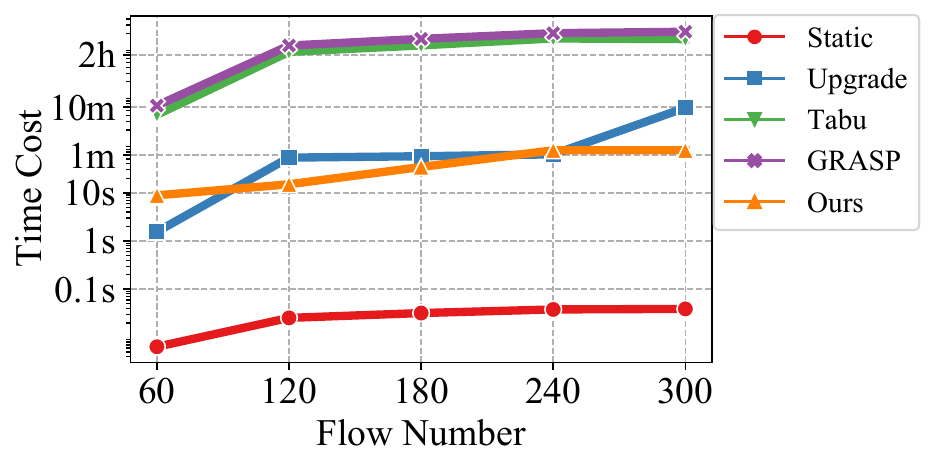}
\caption{Time Cost (log scale).}
\label{fig:time_flownum}
\end{subfigure}
\vspace{-0.5em}
\caption{Performance of algorithms in different traffic volumes.}
\vspace{-1em}
\end{figure*}

\begin{figure*}[t]
\centering
\begin{subfigure}[t]{.27\linewidth}
\centering
\includegraphics[width=\linewidth]{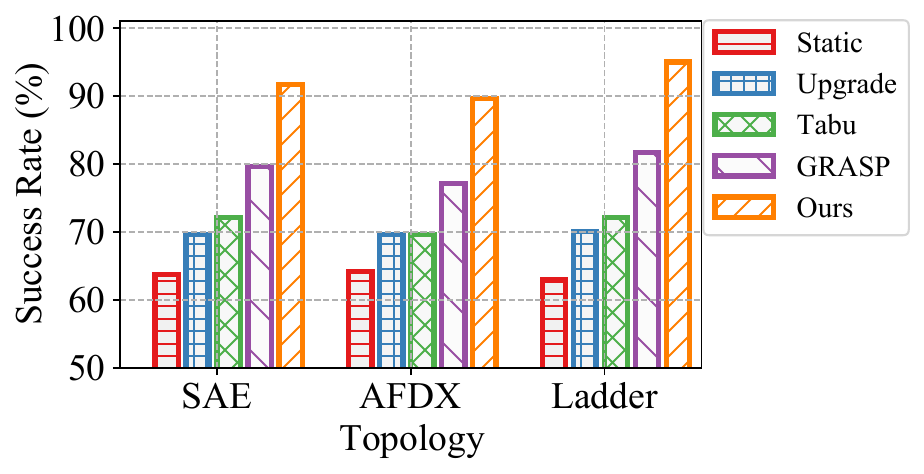}
\caption{Scheduling success rate.}
\label{fig:sucrate_topo}
\end{subfigure}
\hfill
\begin{subfigure}[t]{.27\linewidth}
\centering
\includegraphics[width=\linewidth]{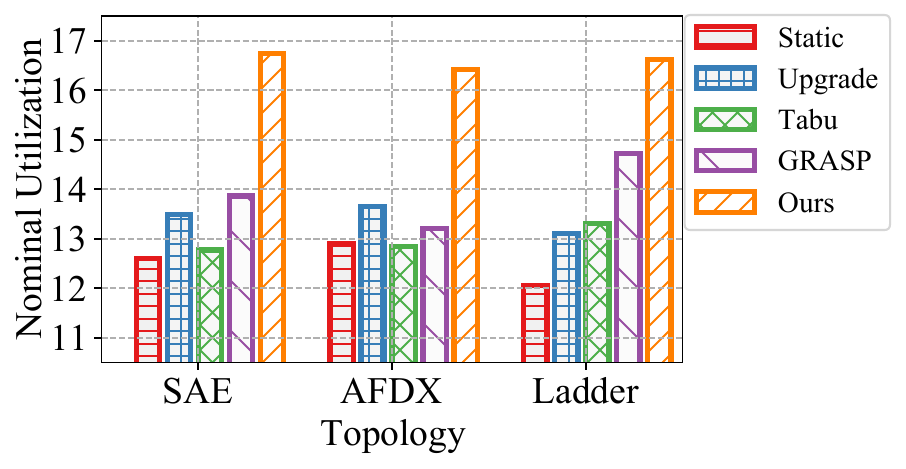}
\caption{Nominal Utilization.}
\label{fig:resuti_topo}
\end{subfigure}
\hfill
\begin{subfigure}[t]{.27\linewidth}
\centering
\includegraphics[width=\linewidth]{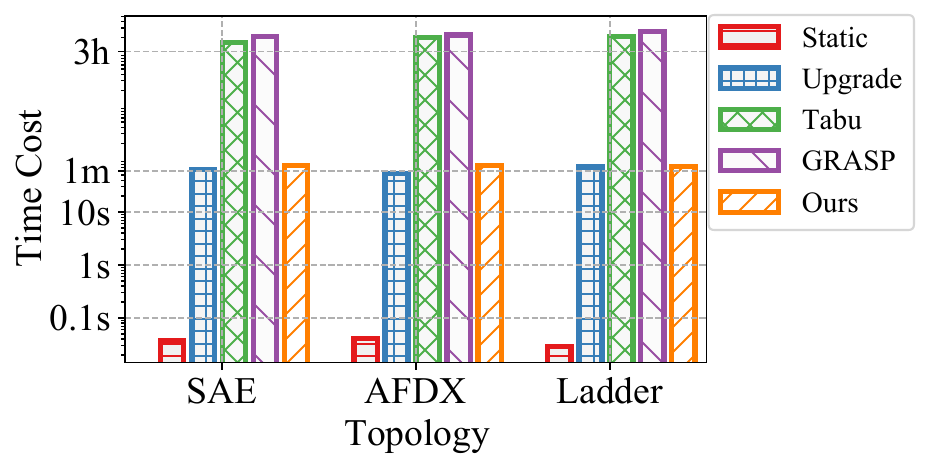}
\caption{Time Cost (log scale).}
\label{fig:time_topo}
\end{subfigure}
\vspace{-0.5em}
\caption{Performance of algorithms in different network topologies.}
\vspace{-1.5em}
\end{figure*}

Fig.~\ref{fig:sucrate_flownum} depicts the schedulability of different algorithms. The average scheduling success rate of the proposed algorithm reaches 92.22\%, exceeding the GRASP algorithm by 15.32\%. It shows the great robustness of the proposed algorithm to different traffic volume patterns. The reason lies in the sorting policies of priority adjustment and schedulability estimation taking both link overlap and flow features into account.

Fig.~\ref{fig:resuti_flownum} shows the nominal utilization of algorithms. It is the total utilization of all types of traffic on all the links\cite{LOBELLO2020153}. It increases with the flow number due to the increasing scheduled flows in the scenario with a large traffic volume. The average nominal utilization of the proposed algorithm exceeds GRASP by 24.12\%, indicating that the link resources are fully utilized by the scheduled flows with the proposed algorithm. 

Fig.~\ref{fig:time_flownum} shows the time costs of different algorithms. The Static algorithm is the least time-consuming because it only conducts scheduling once without any priority adjustment. When the flow number is 300, the time cost of the Upgrade algorithm is 9.6 minutes while the proposed algorithm only takes 1.3 minutes because we separate flows in parallel-FG and cross-FG cases, which can conduct scheduling processes in parallel to save execution time. 
\subsubsection{Performance in different network topologies}

\begin{figure}[t]
\centering
\begin{subfigure}[t]{.24\textwidth}
\centering
\includegraphics[width=\linewidth]{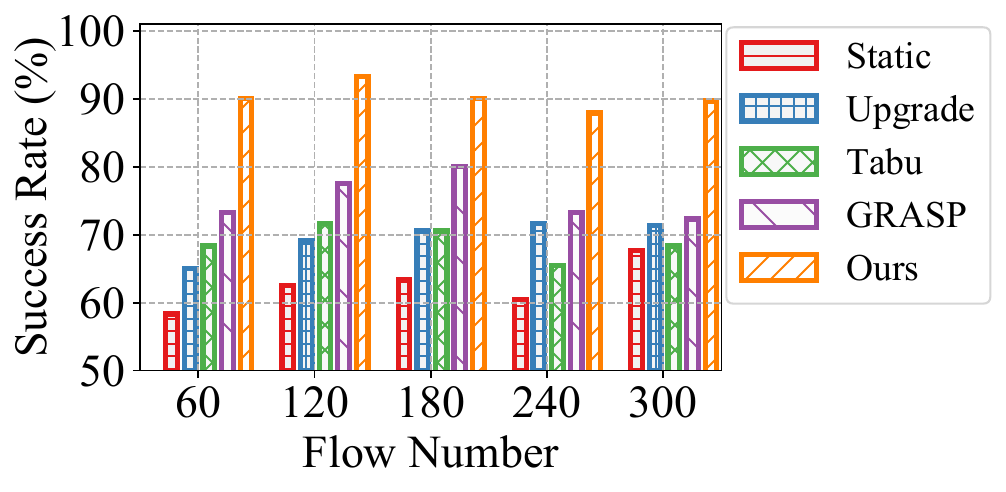}
\caption{Scheduling success rate in parallel-FG case.}
\label{fig:sucrate_para}
\end{subfigure}
\hfill
\begin{subfigure}[t]{.24\textwidth}
\centering
\includegraphics[width=\linewidth]{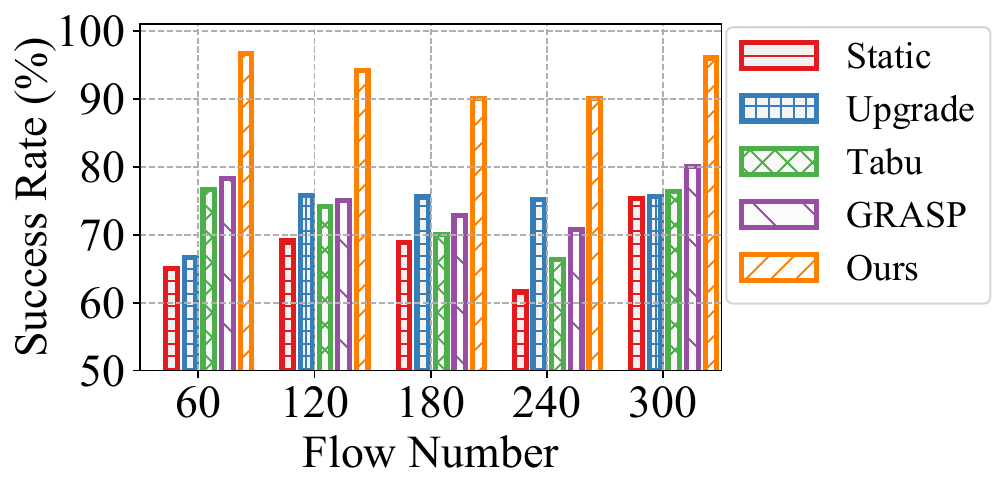}
\caption{Scheduling success rate in cross-FG case.}
\label{fig:sucrate_cross}
\end{subfigure}
\vspace{-0.5em}
\caption{Schedulability of algorithms in separated cases.}
\label{fig:sucrate_para_cross}
\vspace{-1em}
\end{figure}
Fig.~\ref{fig:sucrate_topo} depicts the schedulability of algorithms in different topologies. The average scheduling success rate of the proposed algorithm reaches 92.08\%, exceeding the GRASP, Tabu, Upgrade, and Static by 12.64\%, 20.83\%, 22.36\%, and 28.47\%, respectively. It shows the great robustness of the proposed algorithm to different network conditions. The reason lies in the adaptive consideration of link overlap and flow congestion, and the flexible policy generation in different network scenarios.

Fig.~\ref{fig:resuti_topo} shows the nominal utilization of algorithms. The average nominal utilization of the proposed algorithm exceeds GRASP by 19.10\%, indicating that the link resources are fully utilized by the scheduled flows with the proposed algorithm. The reason lies in its high schedulability.

Fig.~\ref{fig:time_topo} shows the time costs of different algorithms. 
On average, the time cost of Tabu and GRASP are 5.24 hours and 6.27 hours, intolerable in realistic scenarios. The average time cost of the proposed algorithm is 1.23 minutes, striking a balance between scheduling performance and time cost.

\subsubsection{Performance of scheduling in different types of FGs}
To compare the schedulability of algorithms in the parallel/cross-FG cases separately, we run different algorithms in the two scenarios respectively. The results are illustrated in Fig.~\ref{fig:sucrate_para_cross}. In the parallel-FG case, the average success rate of the proposed algorithm reaches 90.18\%, exceeding GRASP, Tabu, Upgrade, and Static by 14.89\%, 21.32\%, 20.64\%, and 27.73\%, respectively. In the cross-FG case, the average success rate of the proposed algorithm is 93.37\%, exceeding GRASP, Tabu, Upgrade, and Static by 17.99\%, 20.65\%, 19.58\%, and 25.37\%, respectively. The above experiments demonstrate the effectiveness of the proposed algorithm in both cases.

\section{Conclusion}
\label{sec:conclusion}
In this work, we introduce the first mixed-criticality traffic scheduling method globally solving the priority adjustment problem among different traffic types with the presence of TAS, CBS, and preemption mechanisms.
We propose adaptive priority adjustment algorithms for flows in different link-overlapping conditions, which can flexibly determine the adjusting policy by exploiting the knowledge of both network conflict and flow features. The experimental evaluations demonstrate the effectiveness of the proposed method.

\bibliographystyle{unsrt}
\bibliography{ref}
\end{document}